\documentclass[12pt]{article}
\usepackage{epsfig,amssymb,amsmath,psfrag}
\usepackage{cite}

\def\IZ{\relax\ifmmode\mathchoice
{\hbox{\cmss Z\kern-.4em Z}}{\hbox{\cmss Z\kern-.4em Z}}
{\lower.4pt\hbox{\cmsss Z\kern-.4em Z}}
{\lower1.2pt\hbox{\cmsss Z\kern-.4em Z}}\else{\cmss Z\kern-.4em Z}\fi}
\newcommand{\Z}{\mathsf{Z}\kern -5pt \mathsf{Z}}
\newcommand{\unit}{\mathsf{1}\kern -3pt \mathsf{l}}

\def\f{\tilde{f}}
\def\Ell{{(L)}}
\def\Ellk{{(L,k)}}
\def\Elltwok{{(L,2k)}}
\def\Elltwokplus{{(L,2k+1)}}
\def\Ellplus{{(L+1)}}
\def\Zero{{(0)}}
\def\One{{(1)}}
\def\Two{{(2)}}
\def\Three{{(3)}}
\def\Four{{(4)}}
\def\cN {  {\cal N}  }
\def\cA {  {\cal A}  }
\def\la {\lambda}
\def\ka {\kappa}

\def \be  {\begin{equation}}
\def \ee  {\end{equation}}
\def \ba  {\begin{eqnarray}}
\def \ea  {\end{eqnarray}}

\def \Tr {\mathop{\rm Tr}\nolimits}

\def\eqn#1{eq.~(\ref{#1})} 
\def\eqns#1#2{eqs.~(\ref{#1}) and~(\ref{#2})}

\newcommand{\nn}{\nonumber}

\begin{document}

\begin{flushright}
BOW-PH-152\\
\end{flushright}
\vspace{3mm}

\begin{center}
{\Large\bf\sf  
All-loop group-theory constraints  \\
for color-ordered SU($N$) gauge-theory amplitudes 
}

\vskip 1.5cm 

Stephen G. Naculich\footnote{ 
Research supported in part by the National Science 
Foundation under Grant No.~PHY07-56518.}

\end{center}

\vskip 0.5cm 

\begin{center}
{\it 
Department of Physics\\
Bowdoin College\\
Brunswick, ME 04011, USA
}

\vspace{5mm}
{\tt naculich@bowdoin.edu}
\end{center}

\vskip 3cm

\begin{abstract}

We derive constraints on the color-ordered amplitudes of 
the $L$-loop four-point function in SU($N$) gauge theories 
that arise solely from the structure of the gauge group.
These constraints generalize well-known group theory relations, 
such as U(1) decoupling identities, to all loop orders.

\end{abstract}

\vfil\break

\section{Introduction}
\setcounter{equation}{0}

An exciting recent development in the study of perturbative amplitudes
is the discovery of color-kinematic duality of gauge theory amplitudes 
at both tree and loop level \cite{Bern:2008qj,Bern:2010ue}.
This duality implies the existence of
constraints on tree-level color-ordered amplitudes,
which were proven in 
refs.~\cite{BjerrumBohr:2009rd,Stieberger:2009hq,Feng:2010my,Chen:2011jxa}.
The BCJ conjecture was also verified through three loops 
for the $\cN=4$ supersymmetric Yang-Mills 
four-\cite{Bern:2010ue,Carrasco:2011hw}
and five-point \cite{Carrasco:2011mn,Bern:2011rj} amplitudes.
(See reviews in refs.~\cite{Carrasco:2011hw,Sondergaard:2011iv},
which also contain references to related work on the subject.)

The BCJ constraints on tree-level color-ordered amplitudes hold 
in addition to various well-known SU($N$) group theory relations,
such as the U(1) decoupling or dual Ward identity \cite{Green:1982sw,Mangano:1990by}
and the Kleiss-Kuijf relations \cite{Kleiss:1988ne}.
Group-theory relations also hold for 
one-loop \cite{Bern:1990ux,Bern:1994zx} and two-loop \cite{Bern:2002tk}
color-ordered amplitudes.
They can be elegantly derived by using an 
alternative color decomposition of the 
amplitude \cite{DelDuca:1999ha,DelDuca:1999rs}.

The purpose of this note is extend the SU($N$) group theory 
relations for four-point amplitudes to all loops.
We develop a recursive procedure to derive
constraints satisfied by any $L$-loop diagram 
(containing only adjoint fields)
obtained by attaching a rung between two external legs 
of an $(L-1)$-loop diagram.
We assume that the most general  $L$-loop color
factor can be obtained from this subset using Jacobi relations,
an assumption that has been proven through $L=4$.
Using this method, we find four independent group-theory
constraints for color-ordered four-point amplitudes at each loop level 
(except for $L=0$ and $L=1$, where there are one
and three constraints respectively).

The color-ordered amplitudes of a gauge theory are
the coefficients of the full amplitude in a basis
using traces of generators in the fundamental representation
of the gauge group.
Color-ordered amplitudes 
have the advantage of being individually gauge-invariant.
Four-point amplitudes of SU($N$) gauge theories can be
expressed in terms of single and double traces \cite{Bern:1990ux}
\ba 
T_1 &=&  \Tr(T^{a_1} T^{a_2} T^{a_3} T^{a_4})
+ \Tr(T^{a_1} T^{a_4} T^{a_3} T^{a_2}),
\qquad\qquad
T_4 =  \Tr(T^{a_1} T^{a_3}) \Tr(T^{a_2} T^{a_4}) , \nn\\
T_2 &=& \Tr(T^{a_1} T^{a_2} T^{a_4} T^{a_3})
+ \Tr(T^{a_1} T^{a_3} T^{a_4} T^{a_2}), 
\qquad\qquad
T_5=  \Tr(T^{a_1} T^{a_4}) \Tr(T^{a_2} T^{a_3})  ,
\qquad\qquad \label{sixdimbasis} \\
 T_3 &=&  \Tr(T^{a_1} T^{a_4} T^{a_2} T^{a_3})
+ \Tr(T^{a_1} T^{a_3} T^{a_2} T^{a_4}), 
\qquad\qquad
T_6 =  \Tr(T^{a_1} T^{a_2}) \Tr(T^{a_3} T^{a_4}) . \nn
\ea
All other possible trace terms vanish in SU($N$) since $\Tr(T^a)=0$.
The color-ordered amplitudes can be further decomposed \cite{Bern:1997nh} 
in powers of $N$ as
\be
\cA^\Ell =
\sum_{\la = 1}^3 
\left( \sum_{k=0}^{\lfloor \frac{L}{2}  \rfloor} N^{L-2k} A^\Elltwok_\la \right) T_\la
+ \sum_{\la = 4}^6 
\left( \sum_{k=0}^{\lfloor \frac{L-1}{2}  \rfloor}N^{L-2k-1} A^\Elltwokplus_\la\right) T_\la
\label{decomp}
\ee
where  $A^{(L,0)}_\la$ are leading-order-in-$N$ (planar) amplitudes,
and  $A^\Ellk_\la$, $k = 1, \cdots, L$,  are subleading-order,
yielding $(3L+3)$ color-ordered amplitudes at $L$ loops.

Alternatively, amplitudes may be decomposed 
into a basis of color factors \cite{DelDuca:1999ha,DelDuca:1999rs}.
It is in such a basis that color-kinematic duality is 
manifest \cite{Bern:2008qj,Bern:2010ue}.
The number of linearly-independent $L$-loop color
factors, however, is less than the number of elements
of the $L$-loop trace basis,
implying the existence of constraints among $A^\Ellk_\la$.
In this note we show that,
for even $L$, the color-ordered amplitudes must satisfy
\ba 
6 \sum_{\la=1}^3 A_\la^{(L,L-2)}  - \sum_{\la=4}^6 A_\la^{(L,L-1)}  &=& 0 \,,
\label{evenrelone}
\\
A_{\la+3}^{(L,L-1)} + A_{\la}^{(L,L)} 
&=& {\rm independent~of~}\la \,,
\label{evenreltwo}
\\
\sum_{\la=1}^3 A_\la^{(L,L)} &=&0 \,,
\label{evenrelthree}
\ea
while for odd $L$, the relations are 
\ba 
6 \sum_{\la=1}^3 A_\la^{(L,L-3)}  - \sum_{\la=4}^6 A_\la^{(L,L-2)}  
+ 2 \sum_{\la=1}^3 A_\la^{(L,L-1)}  
&=& 0 \,,
\label{oddrelone}
\\
6 \sum_{\la=1}^3 A_\la^{(L,L-1)}  -  \sum_{\la=4}^6 A_\la^{(L,L)}  
&=& 0 \,,
\label{oddreltwo}
\\
A_{\la}^{(L,L)} &=& {\rm independent~of~}\la  \,.
\label{oddrelthree}
\ea
These constraints generalize known group theory relations
at tree-level \cite{Green:1982sw,Mangano:1990by},
one loop \cite{Bern:1990ux}, and two loops \cite{Bern:2002tk}
to all loop orders.
In particular, we note that eqs.~(\ref{evenrelone}), 
(\ref{evenrelthree}), (\ref{oddreltwo}),  and
(\ref{oddrelthree}) can alternatively be derived 
by expanding the amplitude in a U($N$) trace basis
and requiring that any amplitude containing
one or more gauge bosons in the U(1) subgroup vanish.
Such U(1) decoupling arguments, however, 
cannot be used to obtain \eqns{evenreltwo}{oddrelone}.

Since the space of $L$-loop color factors is by construction
at least $(3L-1)$-dimensional (for $L \ge 2$),
eqs.~(\ref{evenrelone})-(\ref{oddrelthree}) 
are the maximal set of constraints on color-ordered 
amplitudes that follow from SU($N$) group theory alone.\footnote{If
our recursive procedure together with the Jacobi relations
do not yield the entire space of $L$-loop color factors,
then some of these constraints could be violated for $L>4$,
though we think this unlikely.} 
It is interesting that these constraints only involve the three or
four most-subleading-in-$1/N$ color-ordered amplitudes
at a given loop order;  
other amplitudes are not constrained at all by group theory.
Of course, color-kinematic duality implies further relations
among the amplitudes \cite{Bern:2008qj,Bern:2010ue}.
Other recent work on constraints among loop-level amplitudes 
includes refs.~\cite{BjerrumBohr:2011xe,Feng:2011fja,Boels:2011tp}.

In sec.~\ref{sect-colortrace}, we describe the relation
between color and trace bases, and how to use this
to derive constraints among color-ordered amplitudes.
In sec.~\ref{sect-constraints}, 
we apply this to four-point amplitudes through two loops,
and then develop and solve all-loop-order recursion relations
yielding constraints for four-point color-ordered amplitudes.
In the appendix, we provide details about the three- and four-loop
cases.

\section{Color and trace bases}
\label{sect-colortrace}
\setcounter{equation}{0}

In this section, we schematically outline the approach
we use to obtain constraints among color-ordered amplitudes.
This approach was used in ref.~\cite{Naculich:2011fw} 
for tree-level and one-loop five-point amplitudes.

The $n$-point amplitude in a gauge theory containing only fields
in the adjoint representation  of SU($N$)
(such as pure Yang-Mills or supersymmetric Yang-Mills theory)
can be written in a loop expansion, 
with the $L$-loop contribution given by a sum of $L$-loop 
Feynman diagrams.
Suppressing $n$ and $L$, as well as all
momentum and polarization dependence,
we can express the $L$-loop amplitude in the ``parent-graph''
decomposition \cite{Bern:2010tq}
\be
\cA
= 
\sum_i  a_i c_i
\label{colorbasis}
\ee
where  $\{ c_i \}$ represents a complete set of color factors
of $L$-loop $n$-point diagrams built from cubic vertices
with a factor of the SU($N$) structure constants $\f^{abc}$ at each vertex.    
Contributions from Feynman diagrams containing 
quartic vertices with factors of 
$\f^{abe} \f^{cde}$,  $\f^{ace} \f^{bde}$, and $\f^{ade} \f^{bce}$
can be parceled out among other diagrams containing only cubic
vertices.
The set of color factors may be overcomplete,
in which case they satisfy relations of the form
\be
\sum_i \ell_i c_i = 0.
\label{colorconstraints}
\ee
In fact, it is often necessary to use an overcomplete basis 
to make color-kinematic duality manifest \cite{Bern:2008qj,Carrasco:2011mn}.
Although the amplitude (\ref{colorbasis}) is gauge invariant, 
the individual terms in the sum may not be.
Any gauge-dependent pieces of the form $a_i = \ell_i f$
(where $f$ is independent of $i$)
will cancel out due to \eqn{colorconstraints}.

The $L$-loop amplitude may alternatively be expressed in terms 
of a trace basis $\{  t_\la \}$ as
\be
\cA = \sum_\la  A_\la   t_\la 
\label{tracebasis}
\ee
where $A_\la$ are gauge-invariant color-ordered amplitudes.
One can convert the amplitude (\ref{colorbasis})
into the trace basis by writing 
\be
\f^{abc} =
i \sqrt2 f^{abc} = \Tr( [T^a, T^b] T^c )
\ee
and using the SU($N$) identities
\ba
\Tr(P T^a) \Tr(Q T^a) &=& \Tr (PQ) - {1 \over N} \Tr(P) \Tr(Q)
\nn\\
\Tr(P T^a Q T^a) &=& \Tr (P) \Tr(Q) - {1 \over N} \Tr(P Q)
\ea
to express the color factor $c_i$  as a linear combination of traces
\be
c_i =  \sum_\la M_{ i \la } t_\la \,.
\label{trans}
\ee
The color-ordered amplitudes are then given by 
\be
A_\la = \sum_i  a_i M_{i \la} \,.
\label{colorordered}
\ee
Any constraints (\ref{colorconstraints})
among the color factors correspond to 
left null eigenvectors of the transformation matrix 
\be
\sum_i \ell_i M_{i \la} = 0.
\ee
The transformation matrix will also have a set of right null eigenvectors 
\be
\sum_\la  M_{i \la} r_\la = 0 \,.
\label{null}
\ee
Each right null eigenvector implies a constraint
\be
\sum_\la  A_\la r_\la = 0
\label{rightrelations}
\ee
on the color-ordered amplitudes.

\section{Constraints on color-ordered  four-point amplitudes}
\label{sect-constraints}
\setcounter{equation}{0}

In \eqn{decomp}, 
we decomposed the $L$-loop four-point amplitude 
in terms of the six-dimensional trace basis $\{ T_\la\}$ 
defined in \eqn{sixdimbasis}.
The $1/N$ expansion 
suggests enlarging the trace basis to 
the $(3L+3)$-dimensional basis $\{ t^\Ell_\la \}$:
\ba
t^\Ell_{1+ 6k} &=&   N^{L - 2k} \, T_1 \,,
\qquad\qquad\qquad t^\Ell_{4+6k} =  N^{L - 2k - 1} \, T_4 \,,
\nn \\
t^\Ell_{2+ 6k} &=&  N^{L-2k} \, T_2 \,,
\qquad\qquad\qquad t^\Ell_{5+6k}  = N^{L - 2k - 1}  \, T_5 \,,
\nn \\
t^\Ell_{3+ 6k} &=&  N^{L-2k} \, T_3 \,,
\qquad\qquad\qquad t^\Ell_{6 +6k}=  N^{L - 2k - 1} \, T_6 \,,
\label{expandedbasis}
\ea
in terms of which \eqn{decomp} becomes
\be
\cA^\Ell = \sum_{\la =1}^{3L+3}  A^\Ell_\la t^\Ell_\la,
\qquad
{\rm where}
\qquad
A^\Ell_{\la + 6k}=
\begin{cases}
A^\Elltwok_\la ,    & \la = 1, 2, 3 \,,\\
A^\Elltwokplus_\la,  & \la = 4, 5, 6 \,.
\end{cases}
\label{expandeddecomp}
\ee
The decomposition (\ref{trans}) of color factors $c_i$  into the trace basis
$\{ t^\Ell_\la \}$
shows that the number of independent 
$L$-loop color factors cannot exceed $3L+3$.   
The dimension of the space of color factors is actually less than this, 
being 2-dimensional at tree level, 
3-dimensional at one loop, 
and $(3L-1)$-dimensional for $L \ge 2$
(only proven for $L \le 4$).
As we will illustrate below, this implies the existence 
of right null eigenvectors (\ref{null})
of the transformation matrix $M^\Ell_{i \la}$ 
and corresponding constraints (\ref{rightrelations})
among the color-ordered amplitudes
$A^\Ell_\la$.

At tree level,  
the space of color factors is 
spanned by the $t$-channel exchange diagram
\be
C^\Zero_{st} = \f^{a_1 a_4 b} \f^{a_3 a_2 b}  = t^\Zero_1 - t^\Zero_3 
\label{tchannel}
\ee
and the corresponding $s$-channel exchange diagram
\be
C^\Zero_{ts} = \f^{a_1 a_2 b} \f^{a_3 a_4 b} = t^\Zero_1 - t^\Zero_2  \,.
\label{schannel}
\ee
The $u$-channel diagram is related to these by the Jacobi identity.
With $\{c_1, c_2\}= \{ C^\Zero_{st} , C^\Zero_{ts}\}$,
the transformation matrix (\ref{trans}) and its right null eigenvector 
(\ref{null}) are
\be
M^\Zero _{i \la}  = \begin{pmatrix} 
 1 & 0 &-1 \\ 
 1 &-1 & 0 \\
\end{pmatrix}, \qquad\qquad
r^\Zero = \begin{pmatrix} 1 \\ 1 \\ 1 \end{pmatrix}
\ee
which implies the U(1) decoupling identity
among color-ordered tree amplitudes \cite{Green:1982sw,Mangano:1990by}
\be
A^\Zero_1 + A^\Zero_2 + A^\Zero_3 = 0 \,.
\ee
This is \eqn{evenrelthree} for $L=0$.  

The color factor of the one-loop box diagram 
\be
C^\One_{st} =C^\One_{ts} =
\f^{e a_1 b} \f^{b a_2 c} \f^{c a_3 d} \f^{d a_4 e} 
= t^\One_1 + 2 (t^\One_4 + t^\One_5 + t^\One_6) 
\ee
and its independent permutations $C^\One_{us}$ and $C^\One_{tu}$
span the space of one-loop color factors,
giving 
\be
M^\One _{i \la}  = \begin{pmatrix} 
1 & 0 & 0& &2 & 2 & 2 \\ 
0 & 1 & 0& &2 & 2 & 2 \\ 
0 & 0 & 1& &2 & 2 & 2 \\ 
\end{pmatrix}\,.
\ee
Alternatively, we can choose\footnote{This 
makes sense since we can use the Jacobi identity
to replace the one-loop box diagram
with another box diagram with permutated legs
plus a tree diagram with one of the vertices replaced
by a triangle diagram.
The latter is proportional to a tree diagram since
$ \f^{d a_1 b} \f^{b a_2 c} \f^{c a_3 d}  = N \f^{a_1 a_2 a_3}$.}
for our basis $N C^\Zero_{st}$  and $N C^\Zero_{ts}$, together with
$C^\One_{st}$, to give
\be
M^\One_{i \la}  = \begin{pmatrix} 
 1 & 0 &-1& &0 & 0 & 0 \\ 
 1 &-1 & 0& &0 & 0 & 0 \\
 1 & 0 & 0& &2 & 2 & 2 \\ 
\end{pmatrix} \,.
\ee
In either case, the transformation matrix has three independent right 
null eigenvectors
\be
r^\One = 
\begin{pmatrix} 6 u \\ -u \end{pmatrix}, \quad
\begin{pmatrix} 0 \\ x  \end{pmatrix}, \quad
\begin{pmatrix} 0 \\ y  \end{pmatrix}, \quad 
\qquad {\rm where } \qquad
u \equiv \begin{pmatrix} 1 \\  1 \\  1 \end{pmatrix} \quad
x \equiv \begin{pmatrix} 1 \\ -1 \\  0 \end{pmatrix} \quad 
y \equiv \begin{pmatrix} 0 \\  1 \\ -1 \end{pmatrix} 
\ee 
implying three relations among the one-loop 
color-ordered amplitudes\cite{Bern:1990ux}
\be 
A^\One_4 = A^\One_5 = A^\One_6 = 2 (A^\One_1 + A^\One_2 + A^\One_3)  \,.
\ee 
These are \eqns{oddreltwo}{oddrelthree} for $L=1$.  

At two loops, the ladder and non-planar diagrams\footnote{It can
be easily shown that any other two-loop diagram is related to
these ones by Jacobi relations.}
 yield the color factors 
\ba
C^{(2L)}_{st} 
&=& \f^{e a_1 b} \f^{b a_2 c} \f^{cgd} \f^{dfe} \f^{g a_3 h} \f^{h a_4 f} 
\quad=\quad t^\Two_1 + 6 t^\Two_6 + 2 t^\Two_7 + 2 t^\Two_8 - 4 t^\Two_9  \,,
\\
C^{(2NP)}_{st}
&=& \f^{e a_1 b} \f^{b a_2 c} \f^{cgd} \f^{hfe} \f^{g a_3 h} \f^{d a_4 f} 
\quad=\quad
  - 2 t^\Two_4 - 2 t^\Two_5 + 4 t^\Two_6 + 2 t^\Two_7 + 2 t^\Two_8 - 4 t^\Two_9  \,.
\qquad\qquad
\ea
The non-planar color factors can be expressed in terms 
of the planar ones,
\ba
3 C^{(2NP)}_{st} =
C^{(2L)}_{st} - C^{(2L)}_{ts} - C^{(2L)}_{us} +  C^{(2L)}_{su}  \,,
\ea
and a linear relation exists among the planar color factor and its
permutations,
\be
0 = C^{(2L)}_{st}- C^{(2L)}_{ts}  
  + C^{(2L)}_{us}- C^{(2L)}_{su}  
  + C^{(2L)}_{tu}- C^{(2L)}_{ut}  \,.
\ee
We could therefore choose five of the six permutations of the 
ladder diagram to span the space of two-loop color factors;
alternatively, we can use 
$N^2 C^\Zero_{st}$, $N^2 C^\Zero_{ts}$, and $N C^\One_{st}$,
together with $C^{(2L)}_{st} $ and $C^{(2L)}_{ts} $, to obtain
\be
M^\Two_{i \la}  = \left(\begin{array}{ccccccccccc}
 1 & 0 &-1& &0 & 0 & 0& &0 & 0 & 0\\ 
 1 &-1 & 0& &0 & 0 & 0& &0 & 0 & 0\\
 1 & 0 & 0& &2 & 2 & 2& &0 & 0 & 0\\ 
 1 & 0 & 0& &0 & 0 & 6& &2 & 2 &-4\\ 
 1 & 0 & 0& &0 & 6 & 0& &2 &-4 & 2\\ 
\end{array} \right) \,.
\ee
The two-loop transformation matrix has four independent right null eigenvectors
\be
r^\Two  
=
\begin{pmatrix}  6 u \\ - u \\ 0 \end{pmatrix}, \quad
\begin{pmatrix}    0 \\   x \\ x \end{pmatrix}, \quad
\begin{pmatrix}    0 \\   y \\ y \end{pmatrix}, \quad
\begin{pmatrix}    0 \\   0 \\ u \end{pmatrix}
\label{twoloopnull}
\ee
implying four two-loop group-theory relations \cite{Bern:2002tk}
\ba
0 &=& A^\Two_4 + A^\Two_7 - 2(A^\Two_1 + A^\Two_2 + A^\Two_3) \nn\\
0 &=& A^\Two_5 + A^\Two_8 - 2(A^\Two_1 + A^\Two_2 + A^\Two_3) \nn\\
0 &=& A^\Two_6 + A^\Two_9 - 2(A^\Two_1 + A^\Two_2 + A^\Two_3)  \nn\\
0 &=& A^\Two_7 + A^\Two_8 + A^\Two_9
\ea
equivalent to eqs.~(\ref{evenrelone})-(\ref{evenrelthree})
for $L=2$.

We now employ a recursive procedure to obtain null
eigenvectors for higher-loop  color factors.
An $(L+1)$-loop diagram may be obtained from an $L$-loop diagram
by attaching a rung between two of its external legs, $i$ and $j$. 
This corresponds to contracting its color factor with
$\f^{a_i a'_i b} \f^{b a'_j a_j}$.
Note that if $i$ and $j$ are not adjacent, 
this will convert a planar diagram into a nonplanar diagram.
First consider the effect of this procedure \cite{Glover:2001af}
on the trace basis (\ref{sixdimbasis})
\be 
T_\la \longrightarrow \sum_{\ka=1}^6 G_{\la\ka} T_\ka 
\, , \qquad\qquad {\rm where} \qquad\qquad
G = 
\begin{pmatrix} 
N A & B \\ C & N D
\end{pmatrix}
\ee
with
\ba
A &=& 
\begin{pmatrix}
e_{12}+e_{14} & 0 & 0 \\
0 & e_{12}+e_{13} & 0 \\
0 & 0 & e_{13}+e_{14} \\
\end{pmatrix} \,,
\qquad
B=
\begin{pmatrix}
0 & 2 e_{14}-2 e_{13} & 2 e_{12}-2 e_{13} \\
2 e_{13}-2 e_{14} & 0 & 2 e_{12}-2 e_{14} \\
2 e_{13}-2 e_{12} & 2 e_{14}-2 e_{12} & 0 \\
\end{pmatrix} \,,
\nn\\
C &=&
\begin{pmatrix}
0 & e_{12}-e_{14} & e_{14}-e_{12} \\
e_{12}-e_{13} & 0 & e_{13}-e_{12} \\
e_{14}-e_{13} & e_{13}-e_{14} & 0 \\
\end{pmatrix}\,,
\qquad
D = 
\begin{pmatrix}
2 e_{13} & 0 & 0 \\
0 & 2 e_{14} & 0 \\
0 & 0 & 2 e_{12} \\
\end{pmatrix}\,,
\ea
where the coefficient of $e_{1j}$ corresponds to connecting legs 
$1$ and $j$. 
On the expanded basis (\ref{expandedbasis}),
the same procedure yields
yields 
\be
t^\Ell_\la \to \sum_{\ka=1}^{3L+6} g_{\la\ka} t^\Ellplus_\ka 
\ee
where $g$ is the $(3L+3) \times (3L+6)$ matrix
\be
g = \left( \begin{array}{cccccc}
	A & B & 0 & 0 & 0 & \hdots \\
	0 & D & C & 0 & 0 & \hdots \\
	0 & 0 & A & B & 0 & \hdots \\
	0 & 0 & 0 & D & C & \hdots \\
	\vdots & \vdots & \vdots & \vdots & \vdots & \ddots \\
	\end{array} \right) \,.
\ee
Next, given some $L$-loop diagram with color factor $c_i^\Ell $, 
we can connect two of its external legs with a rung
to obtain an $(L+1)$-loop diagram with color factor
\be
c_i^\Ellplus =  \sum_{\ka=1}^{3L+6}  M^\Ellplus_{ i \ka } t^\Ellplus_\ka
\ee
where 
\be
M^\Ellplus_{ i \ka } 
= \sum_{\la=1}^{3L+3}   M^\Ell_{i  \la} g_{\la \ka} \,,
\qquad\qquad {\rm with} \qquad \qquad 
c_i^\Ell =  \sum_{\la=1}^{3L+3}  M^\Ell_{ i \la } t^\Ell_\la \,.
\ee
Now, suppose we possess a
complete set of $L$-loop color factors $\{ c_i^\Ell \}$
and a maximal set of right null eigenvectors
$\{ r^\Ell_\la \}$ :
\be
\sum_{\la=1}^{3L+3}  M^\Ell_{ i \la } r^\Ell_\la =0\,.
\label{rightnullcondition}
\ee
Then the color factors of all $(L+1)$-loop diagrams obtained 
by connecting two external legs of any $L$-loop diagram
will have a right null eigenvector 
\be
\sum_{\ka=1}^{3L+6}  M^\Ellplus_{ i \ka } r^\Ellplus_\ka = 0 
\ee
provided that $r^\Ellplus_\ka$ satisfies
\be
\sum_{\ka=1}^{3L+6}  g_{\la \ka} r^\Ellplus_\ka = {\rm linear~combination~of~} \{ r^\Ell_\la \} \,.
\label{recursive} 
\ee
We can now solve \eqn{recursive} recursively, 
beginning with the set of $L=2$ right null eigenvectors (\ref{twoloopnull}),
the first case with four independent eigenvectors.
The maximal set of right null eigenvectors satisfying \eqn{recursive} is
\be
\{ r^{(2\ell+1)} \} = 
\begin{pmatrix} \vdots \\ 0 \\ 6 u \\ - u \\ 2 u \\   0 \end{pmatrix} , \quad
\begin{pmatrix} \vdots \\ 0 \\ 0 \\    0 \\ 6 u \\ -  u \end{pmatrix} , \quad
\begin{pmatrix} \vdots \\ 0 \\ 0 \\    0 \\   0 \\    x \end{pmatrix} , \quad
\begin{pmatrix} \vdots \\ 0 \\   0 \\    0 \\   0 \\    y \end{pmatrix} , 
\qquad
\{ r^{(2\ell)} \} = 
\begin{pmatrix} \vdots \\ 0 \\ 6 u \\ - u  \\ 0  \end{pmatrix},  \quad
\begin{pmatrix} \vdots \\ 0 \\   0 \\   x  \\  x \end{pmatrix}, \quad
\begin{pmatrix} \vdots \\ 0 \\   0 \\   y  \\  y \end{pmatrix}, \quad
\begin{pmatrix} \vdots \\ 0 \\   0 \\    0 \\  u \end{pmatrix} \,.
\label{rightnull}
\ee
The constraints on color-ordered amplitudes
\be
\sum_\la  A^\Ell_\la r^\Ell_\la = 0
\label{constraints}
\ee
that follow from the set of right null eigenvectors (\ref{rightnull})
can be written in terms of \eqn{expandeddecomp}
to yield the constraints ~(\ref{evenrelone})-(\ref{oddrelthree})
given in the introduction.

Since there are generally four\footnote{One for $L=0$ and three for $L=1$.}
linearly-independent null eigenvectors in a 
$(3L+3)$-dimensional trace space, 
the space of $L$-loop color factors 
satisfying \eqn{rightnullcondition} 
is generally $(3L-1)$-dimensional.\footnote{Two-dimensional for $L=0$
and three-dimensional for $L=1$.}
Since there are no further independent solutions of \eqn{recursive},
we have shown that the full space of $L$-loop color factors is
{\it at least} $(3L-1)$-dimensional.

We have not strictly shown that \eqn{rightnull} 
are null eigenvectors for {\it any} possible
color factor associated with an $L$-loop diagram,
but rather only for those that can be obtained from an $(L-1)$-loop
diagram by attaching a rung between two external legs.
It is therefore conceivable (but we think unlikely)
that the space of {\it all} $L$-loop 
color factors could be greater than $(3L-1)$-dimensional. 
However, for $L=3$ and $L=4$, it has been shown \cite{Bern:2010tq}
that, despite the fact that many diagrams cannot be
obtained by attaching a rung to the external legs of lower-loop diagrams,
all color factors can be related to these using Jacobi relations  
(see the appendix for further discussion of $L=3$ and $L=4$).
It would be nice to have a proof of this for all $L$, however.

\section{Conclusions}
\setcounter{equation}{0}

In this note, we have extended 
known group theory identities for four-point 
color-ordered amplitudes in SU($N$) gauge theories 
to all loop orders.
We have shown that color-ordered amplitude generally
must satisfy four independent relations at each loop order
(except for $L=0$ and $L=1$, where there are one
and three constraints respectively).
This was achieved via a recursive procedure 
that derives the constraints on $L$-loop color factors 
generated by attaching a rung between two external legs 
of an $(L-1)$-loop color factor.
Assuming that all $L$-loop color factors are linear
combinations of those just described (i.e., via Jacobi 
relations), then the constraints derived 
apply to all $L$-loop color-ordered amplitudes.
Although this has been established through four loops,
it would clearly be desirable to have an all-orders
proof of this assumption.

The recursive method employed in this note 
can also be extended to $n$-point functions with $n>4$
to yield constraints on the color-ordered amplitudes
beyond those already known at tree- \cite{Kleiss:1988ne}
and one-loop \cite{Bern:1990ux,Bern:1994zx} level,
although the size of the color basis grows quickly with $n$.

\section*{Acknowledgments}
It is a pleasure to thank L.~Dixon for valuable correspondence.

\appendix
\section{Appendix}
\setcounter{equation}{0}

In appendix B of ref.~\cite{Bern:2010tq},
bases for the space of all three- and four-loop color factors 
were identified.
In this appendix, we explicitly check that
the right null eigenvectors of these spaces 
coincide with our recursive solution (\ref{rightnull}),
and therefore that all three- and four-loop color-ordered amplitudes
indeed satisfy the group theory constraints
eqs.~(\ref{evenrelone})-(\ref{oddrelthree}).

The basis for three-loop color factors can be chosen as 
$N^3 C^\Zero_{st}$, $N^3 C^\Zero_{ts}$, $N^2 C^\One_{st}$,  
$N C^{(2L)}_{st} $, and $N C^{(2L)}_{ts} $,
plus the color factor for the three-loop ladder diagram
\be
C^{(3L)}_{st} =
t^\Three_1 + 14 t^\Three_6 + 2t^\Three_7 + 2t^\Three_8 
+ 8 t^\Three_{10}+ 8  t^\Three_{11}  + 8 t^\Three_{12}
\ee
and two of its permutations,\footnote{
Only $C^{(3L)}_{st}$ is used in ref.~\cite{Bern:2010tq},
but the authors also include $N C^\Zero_{st}$ and $N C^\Zero_{ts}$
in their basis,
which in our approach are independent of
$N^3 C^\Zero_{st}$ and $N^3 C^\Zero_{ts}$.}
$C^{(3L)}_{ts}$ and $C^{(3L)}_{us}$,
yielding the transformation matrix
\be
M^\Three_{i \la}  = \left( \begin{array}{ccccccccccccccc}
1 & 0 &-1& & 0 & 0 & 0& & 0 & 0 & 0& & 0 & 0 & 0 \\
1 &-1 & 0& & 0 & 0 & 0& & 0 & 0 & 0& & 0 & 0 & 0 \\
1 & 0 & 0& & 2 & 2 & 2& & 0 & 0 & 0& & 0 & 0 & 0 \\
1 & 0 & 0& & 0 & 0 & 6& & 2 & 2 &-4& & 0 & 0 & 0 \\ 
1 & 0 & 0& & 0 & 6 & 0& & 2 &-4 & 2& & 0 & 0 & 0 \\ 
1 & 0 & 0& & 0 & 0 &14& & 2 & 2 & 0& & 8 & 8 & 8 \\
1 & 0 & 0& & 0 &14 & 0& & 2 & 0 & 2& & 8 & 8 & 8 \\
0 & 1 & 0& &14 & 0 & 0& & 0 & 2 & 2& & 8 & 8 & 8 \\
\end{array} \right) \,.
\ee
The four independent right null eigenvectors of this matrix
\be
r^\Three = 
\begin{pmatrix}  6 u \\ - u \\ 2 u \\   0 \end{pmatrix} ,\quad
\begin{pmatrix}   0 \\    0 \\ 6 u \\ -  u \end{pmatrix} ,\quad
\begin{pmatrix}   0 \\    0 \\   0 \\    x \end{pmatrix} ,\quad
\begin{pmatrix}   0 \\    0 \\   0 \\    y \end{pmatrix}
\ee
agree with those in \eqn{rightnull},
and imply the four constraints among the color-ordered amplitudes
given by eqs.~(\ref{oddrelone})-(\ref{oddrelthree}) with $L=3$.

The four-loop color basis can be chosen as 
($N$ times) the three-loop basis
plus three color factors from the four-loop ladder diagram and 
two\footnote{
Only $C^{(4L)}_{st}$ and  $C^{(4L)}_{ts}$ are used
in ref.~\cite{Bern:2010tq},
but the authors also include 
$N C^\One_{st}$,
which in our approach counts as independent from 
$N^3 C^\One_{st}$.}
permutations,
$C^{(4L)}_{st}$, $C^{(4L)}_{ts}$, and $C^{(4L)}_{us}$,
yielding
\be
M^\Four_{i \la}  = \left( \begin{array}{ccccccccccccccccccc}
1 & 0 &-1& & 0 & 0 & 0& & 0 & 0 & 0& & 0 & 0 & 0& & 0 & 0 & 0 \\
1 &-1 & 0& & 0 & 0 & 0& & 0 & 0 & 0& & 0 & 0 & 0& & 0 & 0 & 0 \\
1 & 0 & 0& & 2 & 2 & 2& & 0 & 0 & 0& & 0 & 0 & 0& & 0 & 0 & 0 \\
1 & 0 & 0& & 0 & 0 & 6& & 2 & 2 &-4& & 0 & 0 & 0& & 0 & 0 & 0 \\ 
1 & 0 & 0& & 0 & 6 & 0& & 2 &-4 & 2& & 0 & 0 & 0& & 0 & 0 & 0 \\ 
1 & 0 & 0& & 0 & 0 &14& & 2 & 2 & 0& & 8 & 8 & 8& & 0 & 0 & 0 \\
1 & 0 & 0& & 0 &14 & 0& & 2 & 0 & 2& & 8 & 8 & 8& & 0 & 0 & 0 \\
0 & 1 & 0& &14 & 0 & 0& & 0 & 2 & 2& & 8 & 8 & 8& & 0 & 0 & 0 \\
1 & 0 & 0& & 0 & 0 &30& & 2 & 2 & 0& & 0 & 0 &24& & 8 & 8 &16 \\
1 & 0 & 0& & 0 &30 & 0& & 2 & 0 & 2& & 0 &24 & 0& & 8 &16 & 8 \\
0 & 1 & 0& &30 & 0 & 0& & 0 & 2 & 2& &24 & 0 & 0& &16 & 8 & 8 \\
\end{array} \right) \,.
\ee
The four independent right null eigenvectors of this matrix
\be
r^\Four =
\begin{pmatrix} 0 \\ 0 \\ 6 u \\ - u  \\ 0  \end{pmatrix}, \quad
\begin{pmatrix} 0 \\ 0 \\   0 \\   x  \\  x \end{pmatrix}, \quad
\begin{pmatrix} 0 \\ 0 \\   0 \\   y  \\  y \end{pmatrix}, \quad
\begin{pmatrix} 0 \\ 0 \\   0 \\    0 \\  u \end{pmatrix}, \quad
\ee
agree with those in \eqn{rightnull}.
The right null eigenvalues imply the four relations among 
color-ordered amplitudes
given by eqs.~(\ref{evenrelone})-(\ref{evenrelthree})
for $L=4$.


\end{document}